\documentclass[useAMS]{mn2e}
\usepackage[english]{babel}
\usepackage[latin1]{inputenc}
\usepackage{amssymb}
\usepackage{graphicx}
\usepackage{color}
\usepackage{latexsym}
\usepackage{caption}
\usepackage{cite}
\title{On the Number of Cosmic Strings} 
\author[R. Consiglio et al.]{R. Consiglio$^{1}$ \thanks{E-mail: rosa.consiglio@na.infn.it}, O. Sazhina$^{2}$, G. Longo$^{1,3}$, M. Sazhin$^{2}$, F. 
Pezzella$^{4}$\\
$^{1}$  Dipartimento di Fisica, Universit\`a degli Studi di Napoli ``Federico II'',  Complesso Universitario di Monte S. Angelo ed. 6, \\
\hspace{0.25cm}via Cintia,  80126 -  Napoli, Italy\\
$^{2}$ Sternberg Astronomical Institute,  M.V. Lomonosov Moscow State University, University pr. 13 - Moscow, Russia\\
$^{3}$ Department of Astronomy, California Institute of Technology, Pasadena, 90125 CA, USA\\
$^{4}$ Istituto Nazionale di Fisica Nucleare, Sezione di Napoli,  Complesso Universitario di Monte
S. Angelo ed. 6, via Cintia,  80126 -\\
\hspace{0.25cm}Napoli, Italy}

\begin{document}

\date{Accepted ????. Received ???? in original form ????}

\pagerange{\pageref{firstpage}--\pageref{lastpage}} \pubyear{2002}

\maketitle

\begin{abstract}
The number of cosmic strings in the observable universe is relevant in determining the probability of detecting such cosmic defects through their gravitational signatures. In particular, we refer to the observation of  gravitational lensing events and anisotropy in the CMB radiation induced by cosmic strings. In this paper, a simple method is adopted to obtain an approximate estimate of the number of segments of cosmic strings, crossing the particle horizon, which fall inside the observed part of the universe. We show that there is an appreciable difference in the expected number of segments which differentiates cosmic strings arising in Abelian Higgs and Nambu-Goto models, and that a different choice of setting for the cosmological model can lead to significant differences in the expected number of cosmic string segments. Of this number, the fraction realistically detectable may be considerably  smaller.
\end{abstract}


\section{Introduction}
\label{intro}
Cosmic strings are line-like topological defects which may have formed during a symmetry breaking phase transition in the early universe. The existence of topological defects in spacetime was first theorized by Kibble in 1976 (Kibble 1976). Formation and evolution of cosmic defects were extensively studied in the subsequent decades (Vilenkin \& Shellard 1995; Durrer 1999; Vachaspati 2001) for their cosmological implications. Due to the compatibility of their existence with  cosmological observations and their topological stability, a special interest has always been focused on  cosmic strings (Shellard \&  Allen 1990 ASa hereafter; Hindmarsh \&  Kibble 1994; Hindmarsh, Stuckey \& Bevis 2009). This class of linear defects is a generic prediction in quantum field theory and grand unified theories (GUTs) as well as in string and M-theories. The predicted cosmic strings are extremely thin but very massive objects, characterized by a huge energy and interactions essentially of gravitational nature, whose intensity is measured by the dimensionless tension parameter $G \mu / c^{2}$. The value of this quantity is model dependent; for a string produced at a GUT transition it is of order $10^{-6}$ or $10^{-7}$. Although observations rule out the idea of topological defects as an alternative theory to inflation, the coexistence of cosmological inflation and cosmic strings is apparent in those inflationary scenarios which can  give rise to strings in a phase transition that occurs at the last stages of inflation. This is the case of hybrid inflation models in standard cosmology as well as brane/antibrane inflation models in string cosmology which are of hybrid inflation type. In this framework, a composite family of stable strings may exist in certain classes of models of warped compactifications or large compact extra dimensions which yield string tensions decoupled from Planck scale in the range $10^{-12} \leq G\mu / c^{2}\leq 10^{-6}$ and hence compatible with observational limits  (Arkani-Hamed et al. 1998; Antoniadis et al. 1998; Kachru et al. 2003). Differently from solitonic strings, cosmic superstrings can exist as $F$ and $D$ strings leading to more complicated F-D networks involving the formation of junctions between strings of different tensions (Jones et al 2003; Copeland et al. 2004). Although they can have different properties related to the different context in which they arise, cosmic superstrings are expected to share with the ordinary solitonic strings the same scale-invariance property of the characteristic length scales of the network that they can form. 

Since their energy density may influence the dynamics of the universe, consistency with present day observations imposes constraints on the number density of cosmic strings. To date, cosmic (super)strings are still theoretical objects, nevertheless theoretical advances together with upcoming observational data might provide stronger constraints on their characteristic parameters and make their eventual detection a test of generic predictions of standard or alternative theories. Observations of such strings would provide direct information on fundamental physics and evolution of our universe and maybe the first experimental evidence of a string theory cosmological model underlying the structure of spacetime.

Of the two components of an evolving string network, sub-horizon closed loops and long strings, i.e., infinite strings and large loops with a curvature radius much larger than the horizon size, we are concerned  in particular with the long string component for the effects produced by the presence of this component of the network. Cosmic (super)strings may have a number of cosmological effects. The most interesting observational signatures stem from their gravitational interactions. While specific signatures are expected in models where strings couple to other forces, gravitational effects are common to all cosmic (super)strings and are controlled by the dimensionless tension parameter. A special class of gravitational effects originates from the string peculiar way to deform the spacetime around it. The spacetime around a cosmic string is conical, namely, it is flat everywhere except for a missing wedge where points on the opposite edges are identified. Thus, spacetime looks like a cone with a singularity located at the apex of the cone. Any circular path at a constant distance from the apex around the string is less than a circumference by the deficit angle $\Delta = 8\pi G\mu /c^{2}$. This geometry leads to observable effects, such as the Kaiser-Stebbins effect (KS) (Kaiser \& Stebbins 1984) and gravitational lensing. The former appears as line discontinuities in the CMB temperature caused by a segment of cosmic string moving between the observer and the cosmological photosphere as photons passing by the string move perpendicularly to the plane that contains the string segment. Photons from the last scattering surface streaming by either side of the string will be observed with a redshift or blueshift (ahead of or behind the moving string respectively) proportional to the string tension and the string velocity transverse to the line of sight. Gravitational lensing by a long cosmic string occurs as light from a distant source beyond the string may reach the observer along two different paths, around either side of the string, thus producing a double image of the source with angular separation proportional to the conical deficit angle $8 \pi G \mu /c^{2}$.

The observational signatures of cosmic strings depend on the characteristic parameters, such as string tension and velocity, and on the details of the evolution of the string network which in turn depend on what happens when two strings intersect. A fundamental feature which differentiates a network of cosmic superstrings from a network of ordinary field-theory cosmic strings is their reduced intercommutation probability $P$ which determines the evolution of a string network towards scaling regime. For classical cosmic strings the string tension is directly related to the energy scale of the symmetry breaking and the intercommutation probability is essentially $P = 1$. In fact, even though ultrarelativistic Abelian Higgs strings can pass through each other due to a double intercommutation with the net result that no ends are exchanged (Achucarro \& de Putter 2006), the typical velocities found in simulations for scaling networks of cosmic strings are below the ultrarelativistic threshold, so that the assumption $P = 1$ generally holds. For F-D networks, $P$ can be significantly less than unity, depending on the type of strings and compactification and on the string coupling constant $g_{s}$. As a consequence of the smaller intercommutation probabilities associated with strings of different tensions, in a cosmic superstrings network each type of string may have a different number density. This means that the same fraction of CMB anisotropy can be sourced either by many light strings or by a few heavy ones but it also means that the total amount of strings in the network must be increased by some factor related to the intercommutation probabilities, which implies stronger constraints on tensions. The eventual presence of Y-shaped junctions of two strings joined by a third after intercommutation processes could be detected by cosmic string gravitational lensing or by observation of the KS effect. In the latter case, a different temperature in each of the three patches of sky is expected to be observed, while in the former case the relativistic motion of the binding strings could also lead to an enhancement of the cosmic string  lensing angle by some factor which would be moderate for a typical network motion (Avgoustidis \& Shellard 2005; Henry Tye et al. 2005; Jackson et al. 2005, Pourtsidou et al. 2011).

On the other hand, the $rms$ velocities comparable to the speed of light found in numerical simulations  (Bennett \& Bouchet 1990 BB hereafter; Allen \& Shellard 1990 ASb hereafter; Martins \& Shellard 2005 MS hereafter; Ringeval et al. 2007 RSB hereafter; Bevis et al. 2007; Battye \& Moss 2010 BM hereafter; Blanco-Pillado et al. 2011 BPOS hereafter) could be sufficiently high as to guarantee the detectability of their signatures in the CMB radiation, at least under certain ideal conditions, if such strings do exist and are field theoretical cosmic strings. As for cosmic superstrings, although the $rms$ velocities can be close to the flat spacetime value $c/\sqrt{2} $, the tension parameter associated with the $F$ component, for a string coupling constant of order $10^{-2}$, is smaller than that of the other components, which have values very close to those of ordinary cosmic strings. Such small values of tensions are still unexplored by means of current technology.

The number of cosmic (super)strings contained in the volume of the effective observable universe is also a relevant parameter to determine the probability of observing cosmic strings in gravitational lensing events as well as the density of step-like temperature discontinuities related to the KS effect (Sazhina et al. 2008; Sazhin et al. 2010). More precisely, the significant parameter is the number of string segments which can be considered approximately straight in a  network of random walk  cosmic (super)strings, since these are expected to have a greater probability to leave detectable and unambiguous signals in the observed volume of the universe. In the present paper we consider both cases of field theory cosmic strings and string theory cosmic superstrings, we compare and discuss the results. 

In Section 2 we describe a simple method to obtain an approximate estimate of the number of segments of cosmic strings crossing the observable universe. In particular, for their observational relevance, we consider the volume inside the last scattering surface and inside the sphere of optical sources respectively for KS effect and gravitational lensing observations. The  discourse is broadened to string networks embedded in cosmological backgrounds, variants of the $\Lambda$CDM, where the constant parameter $\Lambda$ is replaced with a dynamical scalar field. For a comparison of results, we consider, for simplicity, spatially flat cosmological models where the cosmic fluid is a mixture of matter and quintessence or a phantom constituent (a time varying component of the energy density of the universe with negative or super-negative pressure, $p < -\rho$) characterized respectively by the ratio of pressure to energy density in the range $-1 \leq w < 0$ and $w < -1$, whereas $w = -1$ coincides with the cosmological constant case. In the above intervals we choose in particular $-1 \leq w_{Q} \leq -0.5$ and $w_{P} = -1.5$ for the quintessence (Q) and phantom (P) components as they are compatible with constraints from large scale structure and CMB observations combined with SNe Ia data. In section 3 we extend the approach of the previous section to address the case of cosmic superstrings. 
In section 4  some features connected with numerical simulations and their underlying models outlined in relation to the scaling property of the string network upon which relies the estimate of the number of string segments. Finally, we discuss our results with particular reference to the range of redshifts in the visible universe where the calculation of the number of string segments can be trusted, taking into account the informations about the scaling regimes provided by numerical simulations. 
\\

Without a loss of generality, in what follows we shall use the cosmological parameters of the $\Lambda$CDM concordance model, which fit the majority of observations (NASA's Lambda):

\begin{equation} \label{eqn:1}
\Omega_{m} \simeq 0.27   \qquad   \Omega_{\Lambda} \simeq 0.73  \qquad  H_{0} \simeq 71 km /s /Mpc.
\end{equation}
\noindent

\section{Number of segments of cosmic string crossing the observable universe}
\label{sec:1}

\subsection{Cosmic strings networks in a $\Lambda$CDM universe}
\label{sec:2}
An approximate estimate of the number of string segments crossing a given volume of the observable universe can be obtained taking into account that Kibble's one-scale model can be conveniently used to study large-scale properties. The single scale $L$, which can be identified with the persistence length or the mean inter-string distance, is defined as the length such that any volume $L^{3}$ will contain on average a string segment of length $L$. By causality, $L$ is bounded by the size of the particle horizon, $d_{PH}$: 

\begin{equation}\label{eqn:2}
L(t) = A_{i} \ d_{PH}(t)
\end{equation}
\noindent
where: 

\begin{equation}\label{eqn:3} 
0 < A_{i} \leq 1   \qquad   i = r, m 
\end{equation}
\noindent
with $r \equiv radiation$ and $m \equiv matter$. The quantity:

\begin{equation}\label{eqn:4} 
A_{i}^{-1} = \frac{d_{PH}}{L}
\end{equation}
\noindent
expresses the scaling property: in the radiation-dominated era (after the transient friction-dominated period) and in matter-dominated era (after the radiation-matter transitional period) the network approaches a scaling regime in which $L$ remains a constant fraction of the particle horizon, as well as the energy in the form of strings remains a constant fraction of the total energy of the universe. Therefore, the above quantity is the result of the efficiency of the various energy-loss processes that a network of strings undergoes. This provides us with an elementary cell, containing a single string segment, whose volume can be calculated being known $A_{i}$, from numerical simulations, and the (physical) particle horizon distance $d_{PH}$: 

\begin{equation}\label{eqn:5} 
d_{PH}(z_{*}) = \frac{c}{H_{0}}\frac{1}{1 + z_{*}} \int^{\infty}_{z_{*}} \frac{dz}{E(z)}  
\end{equation}
\noindent
where:

\begin{equation}\label{eqn:6} 
E(z)= \left[\sum_{i \in I} \Omega_{i} (1+z)^{3(w_{i} + 1)}\right]^{1/2} \qquad I = \left\{\Lambda, k, m, r\right\} 
\end{equation}
\noindent
with $w_{i}$  the equation-of-state parameter for the $i$th fluid. Considering the entire volume of the observable universe, defined by the distance to the particle horizon, circumscribed by a cube having side $2 d_{PH}(z)$  (\ref{eqn:5}), the total number of string segments for all $z$ in the radiation- and matter-dominated eras,

\begin{equation}\label{eqn:7}
N^{r,m}(z) = \frac{\left[2 \ d_{PH}(z)\right]^{3}}{L^{3}(z)} = \frac{8 }{A^{3}_{r,m}}
\end{equation}
 
\noindent
remains constant in each the two cosmological eras of the universe. Now suppose we consider only a fraction of the volume of the observable universe observed at time $t_{*}$. Defining the distance traveled by a photon emitted at time $t < t_{*}$:

\begin{equation}\label{eqn:8}
d(z_{*}, z ) = \frac{c}{H_{0}}\frac{1}{1 + z_{*}}\int^{z}_{z_{*}} \frac{dz^{\prime}}{E(z^{\prime})}, 
\end{equation}

\noindent
where:

\begin{equation}\label{eqn:9}
d(z_{*}, z ) = B(z_{*}, z)  d_{PH}(z_{*})
\end{equation}

\begin{table*}
\caption{Cosmic string parameters (where $v_{r}$ and $v_{m}$ are respectively the $rms$ velocities in the radiation- and matter-dominated era) from high resolution simulations performed from '90s to date. The results obtained by Battye-Moss (BM) refer to both USM modeling NG and AH strings. All the other results but those obtained by Bevis and coauthors (Bevis et al. 2010 hereafter BHKU) refer to NG simulations.}\label{tab:table_1}

\centering
\resizebox*{1.005\textwidth}{!}{
\begin{tabular}{|ccccccccc|}
\hline
\hline   
&&&&&&&& \\                                                
Model par& BB & ASa,b & MS & RSB & BHKU (AH) & BM (NG/AH)& BPOS & $\Delta z$ \\
\hline
\hline
&&&&&&&&\\
$\frac{G\mu}{c^{2}} $ & $4 \times 10^{-6}$  & $ 1.5 \times  10^{-6}$  &  & $7 \times 10^{-7}$ &$1.8 \times 10^{-6}$ & $\left(2.6/6.4\right) \times 10^{-7}$ & &\\
\hline
&&&&&&&&\\
$v_{r}$ & $0.66 c$  &  $0.62 c$  &  $0.63 c$ &          &  $0.5 c$  & $0.65 c$ / $0.4 c$ & $0.63 c$ &\\
\hline
&&&&&&&&\\
$v_{m}$ & $0.61 c$  &  $0.58 c$  &  $0.57 c$ &          &  $0.5 c$  & $0.60 c$ / $0.4 c$ & $0.59 c$ &\\
\hline
&&&&&&&&\\
$A_{r}$ & 0.14      &  0.13      &   0.13    & 0.16     &   0.26   & 0.13/0.35           & 0.15 &\\
\hline
&&&&&&&&\\
$A_{m}$ & 0.18      &  0.17      &   0.20    & 0.19     &   0.29   & 0.21/0.35           & 0.17 &\\
\hline
&&&&&&&&\\
$N(z)$  &5 -- 300   & 6 -- 360   & 4 -- 220 & 4 -- 260  &  1 -- 73  & 3 -- 190/1 -- 40   &6 -- 360& $ (0.5,7]$\\
\hline
&&&&&&&&\\
$N(z)$  &340 -- 960 & 400 -- 1100 & 250 -- 700 & 290 -- 820&80 -- 230& 210 -- 610/46 -- 130&400 -- 1100&  $[8,100]$ \\
\hline
&&&&&&&&\\
$N(z)$ & (1200,2600) &  (1500,3300)   &  (900,3300) &  (1100,1800) & (300,410)  & (780,3300)/170 & (1500,2100) & 1100   \\
\hline 
\end{tabular}
}
\end{table*}

\noindent
with:  

\begin{equation}\label{eqn:10}
0 <  B(z_{*}, z) \leq 1
\end{equation}

\noindent
the fraction of the total number of string segments which fall inside the volume $\left[2 d(z_{*}, z ) \right]^3$ can be computed as:

\begin{equation} \label{eqn:11}
N^{r,m}(z_{*}, z) = \frac{\left[2 \ d(z_{*}, z ) \right]^{3}}{L^{3}} = \frac{8 \ B^{3}(z_{*}, z)}{A^{3}_{r,m}}.  
\end{equation}

\noindent
$B(z_{*}, z) = 1 $ gives the total number of string segments in the particle horizon volume (\ref{eqn:7}) for each cosmological epoch.
\\

The particle horizon delimits the part of the universe beyond which no causal connection exists. However, an observer cannot even see that far, being at any time the view of the universe limited to the largest volume visible in light. Information on matter as well as cosmic strings lying between such visual horizon and the particle horizon could only be obtained through gravitational radiation that they emit, which cannot as yet be detected directly. 
Therefore, in particular, we are concerned with the volume contained inside the spherical surface, named \textit{last scattering surface} (LSS), given by the set of points in space at a distance such that photons emitted at decoupling time reach present-day observers. Then, at any time $t \geq t_{LSS}$, belonging to the matter-dominated era, the distance to any point inside the LSS can be written as the distance traveled by a photon from the emission point, at a time $t \geq t_{LSS}$, to an observer at present time $t_{0}$:

\begin{equation} \label{eqn:12}
d(z) = \frac{c}{H_{0}} \int^{z}_{0} \frac{dz^{\prime}}{\sqrt{\Omega_{m}\left(1+z^{\prime}\right)^{3} + \Omega_{\Lambda}}}. 
\end{equation}

\noindent
Equation (\ref{eqn:12}) defines the effective observable universe, that is, our backward light cone. 
Let us now consider a cube with side equal to the diameter of the spherical volume centered on the present time observer, $2 d(z)$, with:
 
\begin{equation} \label{eqn:13}
z_{0} < z  \leq z_{LSS}
\end{equation}

\noindent
where $z_{0} = 0$ is the redshift at present time and $z_{LSS} = 1100$ is the redshift corresponding to the last scattering time. The largest cube, for $z = z_{LSS}$, circumscribes the CMB sphere so that all strings that enter the last scattering surface and may have potentially observable effects on the CMB radiation are inside this volume. For any given value of the redshift in the interval (\ref{eqn:13}) the number of string segments crossing the corresponding volumes is the fraction of (\ref{eqn:7}) given by:

\begin{equation} \label{eqn:14}
N(z) =  \frac{8 \ B^{3}(z)}{A^{3}_{m}} 
\end{equation}

\noindent
 shown in Table 1 for different simulations NG and AH. 

\subsection{Dark energy cosmological models}
In the previous section we chose for definiteness a particular FRW cosmology, the $\Lambda$CDM model, as the cosmological background for the evolving strings network. In the present section we generalize to a broader class of models by replacing the static homogeneous energy component, with positive energy density and negative pressure $\Lambda$, with a dynamical, time-dependent and spatially inhomogeneous energy component, named \textit{dark energy}, whose negative pressure drives the accelerated expansion of the universe currently observed. In particular, the class of cosmological models we consider are spatially flat FRW space-times dominated by dark energy at late times, after radiation (photons and relativistic neutrinos) and matter (ordinary and cold dark matter) dominance. 
\\

The spacetime, for a spatially-flat expanding universe, is described by the metric:

\begin{equation} \label{eqn:15}
ds^{2} = c^{2}dt^{2} - a^{2}(t)d\textbf{x}^{2}.
\end{equation}
\noindent
Defining the conformal time $\eta$ by:

\begin{equation} \label{eqn:16}
d\eta = \frac{c dt}{a(t)}
\end{equation}
\noindent
the metric can be conveniently expressed in terms of $\eta$ as:

\begin{equation} \label{eqn:17}
ds^{2} = a^{2}(\eta)\left(d\eta^{2} - d\textbf{x}^{2}\right).
\end{equation}
\noindent
The scale factor $a(\eta)$ is determined from the Friedmann equation, which can be written in the form:

\begin{equation} \label{eqn:18}
\left( \frac{d a(\eta)}{a^2(\eta) d\eta} \right)^{2}=H_{0}^{2} \left[ \Omega_m \left(
\frac{1}{a(\eta)} \right)^3 + \Omega_{DE}\left(\frac{1}{a(\eta)} \right)^{3(1+w)} \right]
\end{equation}

\noindent
where $\Omega_{DE}$ is the dark energy density parameter, related to the current contribution of about $73\%$ to the total energy density of the universe, whose budget should also contain the contribution of radiation (as well as spatial curvature), which we have neglected being about one percent. $w = w_{Q}, w_{P}$ is the ratio of pressure to energy density which defines the equation of state for dark energy:

\begin{equation} \label{eqn:19}
p_{DE}=w c^2 \rho_{DE}.
\end{equation}

\noindent 
Current cosmological observations restrict the values of the equation-of-state parameter in the interval: 

\begin{equation} \label{eqn:20}
w \in [-1.5, -0.5].
\end{equation}

\begin{figure*}
\begin{center}
\includegraphics[width=14.5cm]{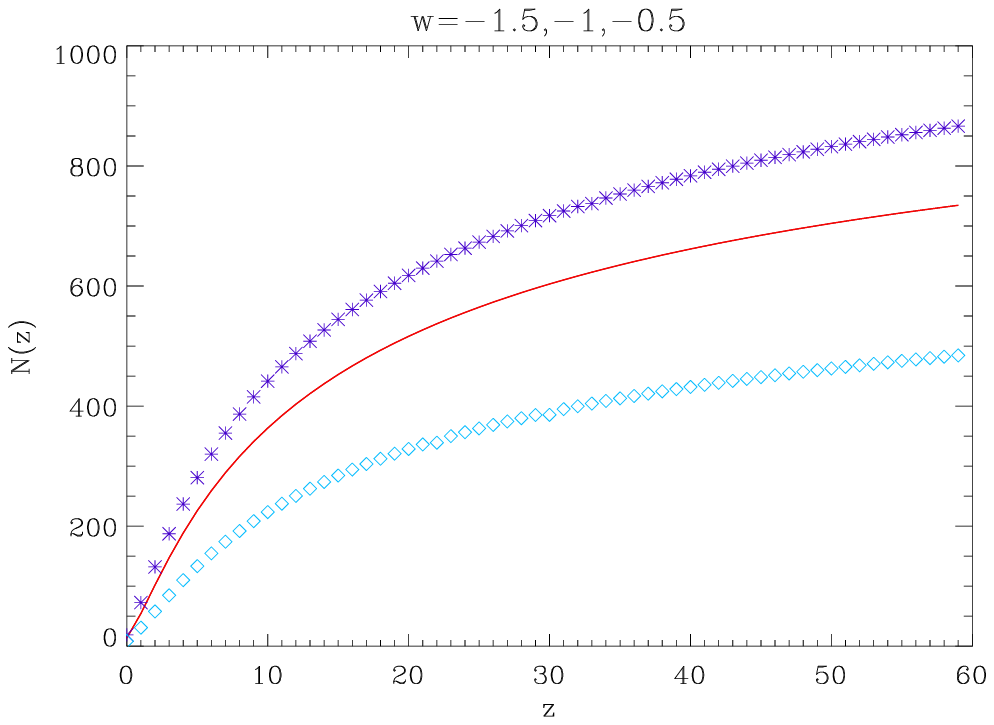}
\end{center}
\caption{$N(z)$ plotted for the equation-of-state parameter varying inside the interval $w \in [-1.5, -0.5]$, for the NG parameters $A_{m,X} \quad X = P, \Lambda, Q$.  The solid  curve corresponds to $w = -1$, stars correspond to $w = -1.5$, while diamonds correspond to $w = -0.5$.}
\label{fig_1}
\end{figure*}

\begin{center}
\begin{table*}
\caption{The approximate number of string segments for different values of the equation-of-state parameter.}\label{tab:table_2}
\resizebox*{0.97\textwidth}{!}{
\begin{tabular}{|cccccccc|}
\hline
\hline  
&&&&&&&\\ 
  Model                              &      P (NG)  &    P (AH)  &  $\Lambda$  (NG)&  $\Lambda$ (AH) & Q (NG)&  Q (AH)   &\\
\hline
&&&&&&&\\ 
w                                       &   --1.5         &  --1.5       &           --1              &        --1               &--0.5      & --0.5    &\\
 \hline
&&&&&&&\\
$A_{m}$                          &    0.18         & 0.30       &       0.19                 &  0.32                   &  0.22      &   0.36   &\\
 \hline
&&&&&&&\\
$A_{r}$                           &    0.17         & 0.29       &       0.18                 &  0.31                   &  0.20      &   0.35   &\\
 \hline
&&&&&&&\\
N(z)                                 &  6 -- 320     &  1 -- 70    & 4 -- 260       & 1 -- 54     &  2 -- 140 &1 -- 33 &\\
$z \in (0.5, 7] $  &&&&&&&\\ 
\hline
&&&&&&&\\
N(z)                               & 360 -- 980    & 78 -- 210    & 290 -- 820       & 61 -- 170     &  160 -- 510&37 --  120 & \\
$z \in[8,100]$    &&&&&&&\\
\hline
&&&&&&&\\
N(z)                               & (1200 -- 1500) & (270 -- 300)  & (1100 -- 1200)  & (220 -- 240)   &  (670 -- 900)&(150 --  170) & \\
$z = 1100$    &&&&&&&\\
\hline
\end{tabular}
}
\end{table*}
\end{center}

\noindent 
Scalar field models can yield a time-varying equation-of-state parameter such that at large redshifts $w$ corresponds to quintessence ($w > -1$), while in later stages, at low redshifts, $w$ becomes much more negative denoting a phantom energy component ($w < -1$). In the very late universe, the phantom energy density ($\rho_{P} > 0$) can grow allowing the parameter $w$ to reach an observationally compatible value slightly below minus one (Caldwell 2002; Copeland et al. 2006; Frieman et al. 2008).

Using the conformal gauge, photons traveling on radial null geodesics, given by $ds^{2} = 0$, move unit comoving distance per unit conformal time $d\eta = \pm dx$. Thus, in the metric conformally flat the light cones are Minkowskian and the speed of light is one. Hence, in conformal coordinates, light emitted at conformal time $\eta_{e}$ will be observed at conformal time $\eta_{o}$ at comoving distance: 

\begin{equation} \label{eqn:21}
D(z_{o}, z_{e}) = \eta_{o} - \eta_{e}. 
\end{equation}

\noindent
Then, the distance to the emitting source at redshift $z_{e}$ can be obtained from the first Friedmann equation as: 

\begin{equation} \label{eqn:22}
D(z_{o}, z_{e}) = \frac{c}{H_{0}} \int\limits_{z_{o}}^{z_{e}} \frac{dz}{(1+z)^{3/2} \sqrt{\Omega_{m}  + \Omega_{DE}(1+z)^{3w}}}.
\end{equation}

\noindent 
In this context the expression of the characteristic length scale of the network is $L_{DE}(t) = A_{m,DE} D_{PH}(t)$. Where  the distance to the particle horizon in terms of $z$ is:

\begin{equation} \label{eqn:23}
D_{PH}(z_{o}) = \frac{c}{H_{0}} \int\limits_{z_{o}}^{\infty} \frac{dz}{(1+z)^{3/2} \sqrt{\Omega_m  + \Omega_{DE}(1+z)^{3w}}},
\end{equation}

\noindent
with $z_{o}  \equiv z_{0} = 0$ for a present-time observer. Using the proper total length in long strings in a particle horizon volume, we can define a quantity that is constant as long as the scaling law holds:

\begin{equation} \label{eqn:24}
\rho_{LS}\ \frac{d^{2}_{PH}}{\mu} = \frac{1}{A^{2}_{r,m}}.
\end{equation}

\noindent 
The long string contribution to the total energy density $\rho$ of the universe scales like radiation in the radiation-dominated era and like matter in the matter-dominated era. Choosing $\Omega_{X} = 0.73$, for $X = \Lambda, Q, P$, $\rho_{LS}$ does not change with the cosmological setting, since it is assumed to be a constant fraction of the same $\rho$. Thus, for quintessence and phantom field models we have:

\begin{equation} \label{eqn:25}
A_{r,m,DE} = \left(\frac{d_{PH,\Lambda}(z)}{d_{PH,DE}(z)}\right)^{3/2} A_{r,m,\Lambda} \qquad DE = Q, P
\end{equation}

\noindent
such that:
 
\begin{equation} \label{eqn:26}
\rho_{LS} =  \frac{\mu}{L^{2}_{X}}    \qquad  X = \Lambda, Q, P
\end{equation}

\noindent
remains the same for each model.

Since the string length density per unit volume $L^{3}$, $1/L^{2}$, is the same for each of cosmological models considered, the number of string segments changes, as we show in Figure 1 and in Table 2. The values of the $A_{m,Q}$ and  $A_{m,P}$ have been derived from the mean values of the  $A_{m,\Lambda}$ related to the NG and AH simulations of Table 1.

\section{Cosmic superstrings}
Cosmic superstrings can form at the end of inflation in brane inflationary scenarios in string theory. They can exist with small tensions, in configurations which are stable  over cosmological time scales and can stretch over cosmological distances as well as solitonic strings. As previously mentioned,  a network of cosmic superstrings  can be in principle distinguished from a network of  field theory cosmic strings by the possibility of forming trilinear vertices connected by a segment of string, when strings of two different types cross, and by the reduced intercommutation probability. The strength of the interaction between the colliding strings depends on the string coupling constant $g_{s}$ and the scale of the confining potential, whose values are not known, and  on the details of compactification, the relative velocity of the strings and their intersection angle.  For the class of models based on large or warped compact extra dimensions, the reconnection probability for the $D$ strings is found to be in the range: $0.1 \lesssim P \lesssim 1$, whereas for the $F$ strings: $10^{-3}  \lesssim P  \lesssim 1 $. When collisions involve strings of different type the reconnection probability can vary from 0 to 1.

New energy-loss mechanisms are also available. As with solitonic strings, oscillating string loops can loose energy through gravitational radiation. In addition, for cosmic superstrings,  massless and massive moduli fields, including the dilaton,  are also expected to be radiated away. Power radiation by $D$-string  loops also includes massless Ramond-Ramond fields. Since the zero modes of the graviton and R-R fields belong to the massless sector of the closed string spectrum, depending on the underlying stringy inflationary model,  these two channels for energy loss might be comparable each other (Firouzjahi 2008). 

\begin{table*}
\centering 
\resizebox*{0.9\textwidth}{!}{
\begin{tabular}{|c|c|c|}
\hline
\multicolumn{3}{c}{$W = 1$  \qquad  $g_{s} = 0.04$} \\
\hline
\hline
&&\\
      $0.6 \lesssim z \lesssim 7$ & $8 \lesssim z \lesssim 100$  & $z = 1100$\\
\hline	
&&\\
      $4000  \lesssim N_{F}(z) \lesssim 2.6 \times 10^{5}$ & $2.9 \times 10^{5} \lesssim N_{F}(z) \lesssim 8.2 \times 10^{5}$  & $1.1 \times 10^{6} \lesssim N_{F}(z) \lesssim 1.8 \times 10^{6}$\\ 
      $2 \lesssim  N_{D}(z) \lesssim 66$ & $73 \lesssim N_{D}(z) \lesssim 210$  & $270 \lesssim  N_{D}(z) \lesssim 330$ \\
      $1 \lesssim N_{FD}(z) \lesssim 35$ & $40 \lesssim N_{FD}(z) \lesssim 110$  &  $140 \lesssim N_{FD}(z) \lesssim 170$ \\
\hline
\end{tabular}
}
\caption{Expected number of string segments of the various components of a F-D network in the small $g_{s}$ regime in two ranges of the redshift, $z \in (0.6,7]$ and $z \in [8,100]$ and at the last scattering epoch.}
\label{tab:table_3}
\end{table*}

\begin{table*}
\centering 
\scalebox{1.2}{
\begin{tabular}{|c|c|c|}
\hline
\multicolumn{3}{c}{\small{$W = 1$ \qquad $g_{s} = 0.9$}} \\
\hline
\hline
&&\\
      $0.6 \lesssim z \lesssim 7$ & $8 \lesssim z \lesssim 100$ & $z = 1100$\\
\hline	
&&\\
      $21 \lesssim N_{F}(z) \lesssim 1300$ & $1500 \lesssim N_{F}(z) \lesssim 4200$ & $5400 \lesssim N_{F}(z) \lesssim 9300$ \\ 
      $5 \lesssim N_{D}(z) \lesssim 300$ & $340 \lesssim N_{D}(z) \lesssim 960$ & $1200 \lesssim  N_{D}(z) \lesssim 1800$  \\
      $0 \lesssim N_{FD}(z) \lesssim 4$ & $5 \lesssim N_{FD}(z)\lesssim 14$&$18 \lesssim N_{FD}(z) \lesssim 20$  \\
\hline
\end{tabular}
}
\caption{Expected number of string segments of the various components of a F-D network in the large $g_{s}$ regime, for $z \in (0.6,7]$ and $z \in [8,100]$ and at the last scattering epoch.}
\label{tab:table_4}		
\end{table*}

Analytical and numerical studies show that although complicated networks may form initially, a scaling regime is eventually reached in which the characteristic lengths are a constant but different fraction of the particle horizon, and that in a scaling superstring network the three lightest strings,  (1, 0), (0, 1) and (1, 1) (respectively $F$ and $D$ strings and the lightest $FD$  bound state) dominate the string number density, whatever the value of $g_{s}$.

For a large value of the string coupling, $g_{s} \sim O(1)$, most of the network energy density is in the lightest (1, 0) and (0, 1), whose tensions in this case are approximately equal being $\mu_{D} = \mu_{F}/g_{s}$. At smaller values of $g_{s} \sim O  (10^{-2})$, the (1, 0) component becomes much lighter than both the (0, 1) and (1, 1) strings and dominates the string number density although, because of their much larger tension, the energy density of the network can be dominated by the less numerous components (0, 1) and (1, 1). 

Being much heavier than the $F$ strings, the $D$ strings may evolve independently of the $F$ strings and reach the scaling regime like a single $D$-strings network. After the $D$ strings settle down the scaling distribution on a given length scale, the $F$ strings in turn may evolve like a single $F$-string network.

Allowing for specific details that depend on the model-dependent value of the effective volume of the compact dimensions, in each of the two scaling behaviors, at large or small values of  $g_{s}$, the energy density of the multi-tension network appears to be dominated by nearly single tension strings. 

However, these two distinct scaling behaviors predicted for a network of cosmic superstrings in principle can be seen in the power spectra of the CMB anisotropies they source, which depend on the type of strings involved. At different values of  $g_{s}$  strings with different correlation lengths can dominate the spectra and determine the position of the main peak.
In the one-scale model, the characteristic length  scale $L(t)$ of the network represents both the correlation length in directions along the string and the average inter-strings distance, supposed to be approximately equal. Thus string networks with higher (lower) number density have smaller (larger) correlation lengths. The values of the correlation length and the $rms$ velocity (both dependent on  $g_{s}$) determine the shapes of the string-induced CMB spectra.

The extension of the velocity dependent one-scale model to a multi-tension F-D network requires defining a specific length scale for any type of string $L_{i} (t)= A_{r/m,i} \ d_{PH}(t)$ ($ i = F, D , FD,$) and the energy density for any component of the network $\rho_{i} = \mu_{i}/ L^{2}_{i}$.

The tension of each component in flat spacetime is given by:

\begin{equation} \label{eqn:27}
\mu_{i} \equiv \mu_{(p_{i},q_{i})} =  \frac{\mu_{F}}{g_{s}} \sqrt{p^{2}_{i}g^{2}_{s} + q^{2}_{i}}
\end{equation}

\noindent
where $\mu_{F} = \mu_{(1,0)}$ and $\mu_{D} = \mu_{(0,1)}$ are respectively the tensions of the lightest $F$ string carrying charge $(1,0)$ and $D$ string carrying charge $(0,1)$, as well as $\mu_{FD}$ becomes essentially $\mu_{(1,1)} = (\mu_{F}/ g_{s}) \sqrt{g^{2}_{s} + 1}$, as the lightest bound state only appears to dominate the string number density.

Following the results obtained in (Pourtsidou et al. 2011, see for more details on the model)  in the two limiting values of the coupling $g_{s} = 0.04$ and $g_{s} = 0.9$ and a particular value of the volume factor associated with compactification $W = 1$, corresponding to a compactification close to the string scale, we find for each component of the network the results, represented in the tables 3 and 4, for the string number in the various ranges of redshift observationally interesting.  

In both $g_{s}$ regimes  the  $F$ component of the network gives a large contribution to the total number of string segments in a  volume corresponding to a given $z$:

\begin{equation} \label{eqn:28}
N(z) = \sum^{3}_{i = 1}N_{i}(z).
\end{equation}
\noindent
The results represented in Table 3 show that for $g_{s} = 0.04$ the expected number of string segments for the $F$ component appears to be exceedingly large. In fact, at small redshifts, down to $z = 0.6$, the order of magnitude of $N(z)$ is $10^{3}$ against a value of about $20$ obtained for $g_{s} = 0.9$. This, on the one hand, suggests to look at an intermediate regime to find hopefully realistic results. On the other hand, the tension parameter associated with the $F$ component, for $g_{s} = 0.04$,  is smaller than that of the other components, which have values very close to those of ordinary cosmic strings. Such small ranges of tensions are still unexplored by means of current technology. In fact, the CMB spectra sourced by a F-D network normalized to give that cosmic strings cannot contribute more than $10\%$ of the total CMB temperature anisotropy, correspond to:

\begin{equation} \label{eqn:29}
\frac{G\mu_{F}}{c^{2}} = 1.8 \times 10^{-8},   \qquad 
\frac{G\mu_{F}}{c^{2}} = 2.1 \times 10^{-7}.
\end{equation}

\noindent
respectively for $g_{s} = 0.04$ and $g_{s} = 0.9$.
 
Despite the higher values of the $rms$ velocities that the $F$ component may reach  in the radiation era $v_{r,F} \sim 0.7 c$, with very  small decrease in the matter era in the small $g_{s}$ regime, for $W = 1$, a tension parameter of order $10^{-8}$ results in a value of the amplitude of the temperature anisotropy of order $10^{-7}$, which is still out of reach, being the order of magnitude of the Plank resolution in temperature $10^{-6}$. 

In the large $g_{s}$ regime the $rms$ velocity of the $F$ component is smaller ($v_{r,F} \sim 0.67 c$) with more evident decrease in the matter era, but the tension parameter (\ref{eqn:29}) is the same order of magnitude of field theory cosmic strings. The length scale parameter is also comparable to that of ordinary cosmic strings.
\\

The one-scale model is unable to provide a proper description of the network properties at scales close to the string width, which is set by the scale of the symmetry breaking phase transition in field theory and  by the fundamental string scale, warp factors associated to the spacetime curvature and size of the internal six-dimensional manifold in string theory. On the other hand, this model well describes not only the large-scale features of the scaling network but also the deviation from scaling. 

\section{Discussion and results}
\subsection{General framework}
The number of string segments $N(z)$ has been determined in the range of redshifts of interest for observations by using the constant quantity $A_{m}$ associated with the scaling law, thus assuming that the network has already reached the scaling regime in the matter-dominated era by the last scattering  time which is the upper limit of the range of redshifts (\ref{eqn:13}). As a matter of fact, there is a transition period between the radiation-dominated and matter-dominated eras where the string network will be far from the type of equilibrium that characterizes a scaling solution for a time dependent on how fast is the network's response to the change in expansion rate. Thus, the range of redshifts, in the matter-dominated era, within which we can trust a string segment distribution based on the scaling property, could be well below $z_{LSS}$ and hence below the range of radio surveys for searching of cosmic string signatures in the CMB radiation. A similar limit, this time at very small redshifts, namely $z \approx 0.5$, is also to be taken into account since there is a further transition between the matter-dominated era and the dark-energy-dominated era (Riess et al. 2004; Bolotin et al. 2012). The effect of this transition to a cosmological era where the universe expands exponentially will be a reduction of the string velocity, which prevents reconnection, affecting the scaling behavior and, at the same time, a dilution which might lead to a small number density of the leftover network (Albrecht et al. 1998; Battye et al. 1998). 

The scaling property, which implies it makes sense to consider the possibility of the  string existence in our universe and search for their observable effects, is the first fundamental information about a network of cosmic strings yielded by numerical simulations. For classical cosmic strings, evolving in a $\Lambda$CDM cosmological background, these simulations are essentially based on the field theory Abelian Higgs (AH) model and the Nambu-Goto (NG) model. The latter relies on the assumption that, as the scale of their diameter is much smaller than any cosmological scale, cosmic line-like defects can be studied in the zero width approximation. 

In particular, in NG simulations the time required for the long string component relaxation to the scaling regime in the matter-dominated era can appear decidedly longer than the transient taken to reach scaling in the radiation-dominated era so that the transition period might last far beyond the end of the decoupling time. Some difference can be found in the NG simulations results related to a different choice of the long component of the network, which can include all loops but those evolving on non-intersecting trajectories (stable loops that cannot rejoin the network) (BPOS) or only super-horizon loops. In the latter case, the long string component could reach the scaling regime in a relatively short time showing little time delay between the change in the expansion law and the response of the network. However, for what it concerns the loop component, its energy density also reaches a scaling evolution but the loop distribution takes more time to reach the scaling regime and the transition period is longer for the smaller loops (BB; RSB).

In AH simulations the effects of matter-radiation transition are not apparent in the network evolution. The first simulations yield approximately even the same value $A_{r} \simeq A_{m} \approx 0.3$ (Bevis et al. 2007), although this result is admittedly a consequence of a simulation size about the minimum required to study a scaling network of strings and recent improved simulations performed by the same authors yield a different result for the scaling parameters in  radiation- and matter-dominated era, with $A_{r} \simeq 0.26$ and $A_{m} \simeq 0.29$ (Bevis et al. 2010). 

The reason for differences between simulation results lies mainly in the model at the basis of the simulations. The AH model includes the small-scale physics near the string width, which has a significant role in the string dynamics. While in AH simulations energy from the strings is converted into massive gauge and Higgs radiation through oscillating short-lived loops as well as  direct emission, in NG simulations this decay channel is not available and the string length density is significantly higher, with the long strings first converted into small loops and subsequently decaying via gravitational radiation. This additional decay channel in the field theoretic simulations by direct massive radiation may be the primary energy-loss mechanism for long strings and cause AH strings to scale without gravitational radiation. In the NG scenario, particles are produced only near cusps, which determines much weaker constraints on the production of extremely energetic Standard Model particles (the Ultra-High Energy Cosmic Rays) than in the AH model. On the other hand, in the AH model there are no constraints from gravitational radiation whereas in the NG case the strongest limits come from millisecond pulsar timing which depend sensitively on details of the loop distribution and dynamics. The number of loops in the simulation volume is substantially less in field theory simulations than in NG where a population of non-intersecting loops remaining stable is found while in AH, where massive radiation is available, loops radiate and shrink. As regards the relevant component for CMB, the long  strings, the results yielded by the two types of simulations are quite different in most respects, such as the inter-string separation and therefore the string density which results significantly lower in AH than in NG simulations. Similarly, the $rms$ velocities probably due to backreaction from the massive radiation are lower in AH than in NG simulations. In fact, AH simulations give $\approx 0.5 c$ (Hindmarsh et al. 2008) in both the radiation and matter eras whereas NG simulations give distinct values in the two cosmological eras both generally above $0.6 c$. 

A specific problem related to AH field theory simulations is that computational constraints require the string width to be artificially increased in order to keep it above the simulation resolution. Although this does not significantly affect the CMB power spectra results a much relevant problem associated with the string width is that the 3D boxes required imply a reduction of the dynamical range. The fact that in AH simulations most of the network energy is emitted through field radiation, until the point that loop production is almost insignificant, has been supposed to be a consequence of a low resolution in AH simulation compared to NG ones. On the other hand, the accuracy of the NG simulations with regard to small-scale structure and loop production has also been questioned. In fact, the NG approximation fails close to the string width, that is, at cusps and kinks which are expected to form in an intersecting string network.
 Furthermore, once energy is transferred via intersection events from long strings to small loops, these are assumed to decay into gravitational radiation and are removed from the simulation since the gravitational waves are not included in any simulation for obvious reasons of complexity connected to backreaction on the string network.

It is worth noting that when simulations involve cosmic superstrings there is none of the aforementioned problems concerning the zero-width approximation. Although cosmic superstrings, as well as conventional cosmic strings, have a tiny but non zero width, they are theoretically one-dimensional objects.  The rich variety of energy loss channels associated with specific properties of such superstrings of cosmic size appears to lead to a rapid approach to the scaling regime for each component of the network. 
\\

An alternative approach to field theory cosmic strings consists in using simulations of the NG or AH type and the \textit{Unconnected Segment Model} (USM). The USM model represents the string network as a stochastic ensemble of unconnected moving segments, of length $A d_{PH}(t)$ and $rms$ velocity $v$, which are randomly removed at an appropriate rate so as to find sub-horizon decay and the proper string scaling density. The USM model with parameters measured in simulations has been used in the context of CMB anisotropies to reproduce the string power spectra coming from different simulation techniques and derive upper bounds on the string tension parameter (BM). This approach also includes the radiation-matter transition and allows, in principle,  to study the effects of deviations from scaling at the onset of a $\Lambda$-dominated phase. 
\\

As regards the effects on CMB of the network evolution during radiation-matter transition, which sets an upper bound to the applicability of eq. (\ref{eqn:14}) at redshifts presumably even well below $z_{LSS} = 1100$, until about $z \sim 100$, it has been shown (Allen et al. 1996) that the large-scale anisotropies are primarily produced at $z \lesssim 100$. Since the coherence length of the string network grows with time, anisotropies on small angular scales are expected to be seeded at early times, and the large angle anisotropies to be seeded at late times. Thus, the contribution of cosmic strings in the redshift range $100 < z < 1100$ should only slightly affect the large angle CMB anisotropy so that the range of interest (\ref{eqn:13}) in this case can be restricted to $z_{0} < z \lesssim 100$. At small angular scales the relevant contribution to temperature anisotropy induced by cosmic strings comes from the last scattering epoch where one may expect the strings' signatures in the CMB temperature fluctuations to be dominated by their \textit{integrated Sachs-Wolfe} effect from the last scattering surface. It has been shown that at high multipoles the mean angular power spectrum of string-induced CMB temperature anisotropies can be described by a power law slowly decaying as $\ell^{-p}$, with $p = 0.889^{+0.001}_{-0.090}$ (Fraisse et al. 2008). This implies that the power spectrum at small angular scales decays much slower than the exponential Silk diffusion damping of the CMB primordial anisotropies. Consequently, the cosmic string contribution to the angular power spectrum can be subdominant at low $\ell$ while it dominates the primary anisotropies for large values of $\ell$. The exponential suppression of the inflationary contribution at high $\ell$ also means that the string component may dominate for a realistic contribution (i. e. with normalization such that string contribution $f_{10}$ at multipole $\ell = 10$ is $10\%$ of the observed power at this multipole) at $\ell \gtrsim 3500$. The fraction of the total theoretical spectrum due to strings $f_{\ell}$ is found to increase (BHKU) from $f_{1500} \approx 0.1$ to $f_{3500} = 0.5$,  which points out that a non negligible cosmic strings contribution to the CMB anisotropies is to be searched at small angular scales, corresponding to $\ell \gtrsim 2000$, as soon as they  will be accessible. Even though these are the same angular scales for which the \textit{Sunyaev-Zel'dovich} effect begins to make a significant contribution to the temperature power spectrum, this occurs at two observational frequency bands, namely, 100 and 150 GHz, whereas the string contribution is at frequencies in observational bands near 220 GHz, where the Sunyaev-Zel'dovich effect is suppressed. 

Although the relaxation time for the cosmic string loops to reach a self-similar evolution with respect to the horizon size appears to be larger for smaller loops, the considerable change in scale factor between the GUT redshift and the last scattering surface might well lead the observable length scales of the infinite component of the network to be in scaling at decoupling, $z = 1089$. Such scaling long strings may contribute to the CMB anisotropies down to fairly small angles. On the other hand, including the non-scaling loops, field theory numerical simulations show some extra power at very small scales, which suggests that non-scaling structures start to have significant effects at very small scales, for $\ell \lesssim10^{4}$. As regards the relevant contribution due to the infinite component to the CMB anisotropies at moderately high $\ell$, the number of string segments in the corresponding range of redshifts close to $z_{LSS}$, where the network is not expected to have reached the matter era scaling regime, can be estimated as:

\begin{equation} \label{eqn:30}
\frac{8 \ B^{3}(z)}{A^{3}_{m}} <  N(z) < \frac{8 \ B^{3}(z)}{A^{3}_{r}}.
\end{equation}

\noindent
Taking into account that it is not known how long it lasts the deviation from scaling, $N(z)$ is expected to be closer to the lower or to the upper bound, depending on how fast the network enters the scaling regime. For cosmic superstrings,  the above relation  (\ref{eqn:30}) applies to each component of the network; numerical simulations (Pourtsidou et al. 2011)  show that  the value of $A_{i,j}$ appears generally much closer to $A_{r,j}$ than it is to $A_{m,j}$ so that $N_{j}(Z)$ should be near the upper bound. 

Let us now consider the deviation from scaling during the matter-dark energy transition, which sets a lower limit at a redshift $z_{0} \approx 0.5$ to the applicability of eq. (\ref{eqn:14}). The number of cosmic string segments at $z \lesssim 1$ could be extremely small so that it would become important to understand how deviation from the scaling behavior around the present cosmological era may affect the network since an eventual network no longer scaling with matter, exponentially slowed down, could be quite difficult to detect. As previously mentioned, the string velocity plays an important role for detecting signatures in the CMB since it determines the amplitude of the line-like discontinuities in the CMB temperature. As for the gravitational lensing there is no such restriction to relativistic velocities for observation to succeed but the number of string segments in the range of optical sources could be a very small fraction of the initial number and approaching $z \lesssim 1$ it could become a small fraction even of the expected number in the actually observable universe. This can be clearly seen from the values of $N(z)$ for $z \in (0.5,7]$ presented in Table 2 for field theory cosmic strings and in Tables 3 and 4 for string theory cosmic superstrings.

Thus, assuming that the most distant objects we are able to observe by optical methods are located at a redshift $z = 7$, the probability of observing segments of the strings crossing the effective observable universe in the optical range is proportional to the volume of the optical sphere, having radius $d(z = 7) = B(z = 7) d_{PH}$. In horizon units then, in a $\Lambda$CDM cosmological background, we have $p \sim B^{3}(7) = 0.22$, and the present angular size of a string having center at this redshift is $\sim 100^{o}$ (Sazhina et al. 2008), whereas the angular size of a segment of length $L(z = 7)$ eventually observed is about $7^{o}$. For smaller values of the redshift, the number of string segments inside the corresponding sphere becomes increasingly smaller and the angular size of the string, with center at that redshift, larger. For instance, at a redshift $z = 2$ the probability is reduced to $p \sim B^{3}(2) = 0.05$, the current angular size of the string is around $134^{o}$ while for a segment $L(z = 2)$ the angular size we should observe is about $20 ^{o}$. Below $z \lesssim 1$ the number of string segments inside the sphere of radius $d(z \lesssim 1)$ is expected to be so small as to render detection by optical methods in this range a challenging problem unless the deviation from scaling at transition to the dark energy era is such that the decrease in reconnection events, whose effect is to increase the density of strings and the time needed to reach equilibrium, may somewhat compensate the eventual dilution by exponential expansion maintaining the number of strings to a detectable level.

Regarding the dark energy variants of the standard cosmology, a different choice of the equation-of-state parameter implies increasing or decreasing distances and hence volumes defined through such distances. These volume effects are due to a different expansion rate: phantom models have always a larger volume than the $\Lambda$CDM model since the expansion rate is larger, while quintessence models have a smaller volume than the $\Lambda$CDM model, being the expansion rate lower. As a consequence, one can only find an increased or decreased number  of string segments, being the characteristic length scale of the network unaltered. The number of string segments for $w \in \left\{-1.5, -1, -0.5\right\}$ is presented in Table 2. 

As we can see from Figure 1, $N(z)$ is strongly dependent on $w$. The graph refers to NG strings but the trend is the same, decreased by a factor of about 4, for AH strings. A comparison of the results obtained in the two ranges of the optical sources, up to $z = 7$,  and radio surveys, which covers all $z$ up to last scattering time,  points out that the distribution of string segments could be sufficient to leave detectable signals through gravitational lensing and in the CMB radiation for NG  strings, whatever the expansion rate. 
\\

The density of step-like temperature discontinuities in the CMB radiation (KS effect) produced by cosmic strings crossing the last scattering surface is determined by the number density of string segments in that volume. On the other hand, deviations from the ideal condition on the string orientation $\hat{k} \cdot  (\hat{\beta_{s}}\times \hat{s}) = 1$ (where $\hat{k}$, $\hat{s}$ and $\hat{\beta_{s}}$  are respectively the unit vectors along the line of sight, the string and its direction of motion), leading to:
\begin{equation} \label{eqn:31}
\frac{\delta T}{T} = \Delta  \gamma_{s} \beta_{s}
\end{equation}
\noindent
(where  $\gamma_{s}$ is the Lorentz factor for the string segment) imply that not all of the string segments distributed in the last scattering surface may leave detectable signatures on the CMB radiation and that in the most favorable case the amplitude $\delta T /T$ of the  temperature fluctuations would be an order of magnitude lower than the adiabatic fluctuations ($\sim 10^{-5}$). In fact, the two positive factors less than one, which appear at the right side of equation (\ref{eqn:31}) when $\hat{\beta_{s}}$, the string direction $\hat{s}$ and the line of sight are not mutually perpendicular, could render very difficult to have a statistics of signals  significant and identifiable unambiguously as sourced by such cosmic strings, especially if the underlying cosmological model is a quintessence model and the string evolution is described by the AH model.

 Regarding cosmic superstrings, the small value of the deficit angle that turns out for the $F$ component in the  limit $g_{s}\sim 10^{-2}$,  $\Delta \sim 0.09^{\prime \prime}$, leads  to temperature jumps in the CMB radiation no longer appreciable so that the presence of a large number of $F$ strings does not conflict with observations. As for the $D$ component of the network, although the temperature discontinuities span over the wide volume of the universe enclosed in the last-scattering surface, the number of $D$ string segments up to $z = 1100$ remains quite small and still smaller is expected to be the number of bound states. Consequently, the probability of observing  peculiar temperature patterns due to Y-junctions appears to be very low, especially when observations are  performed  with ground-based telescopes having a limited field of view covering only a part of the sky. 
For similar reasons, in the range of $z$ of interest for gravitational lensing effects, as we shall see in Section 4.2,  the observation might be quite challenging, especially if cosmic strings are not perpendicular to the line of sight.

\subsection{On the observation of gravitational lensing by cosmic strings }
The difference between the two classical models describing the evolution of cosmic strings is quite significant and it is almost of the same amount for the different cosmological models describing the background in which they evolve. Therefore, for the considerations that follow, for definiteness, as regards field theory cosmic strings, we mainly refer to the $\Lambda$CDM  cosmology, but the extension to the above mentioned variants and to superstring theory cosmic strings is discussed as well and summarized in Table 5.
\\

Although the constraint on the string tension parameter set by gravitational lensing is $G\mu /c^{2} < 3 \times 10^{-7}$ (J. L. Christiansen et al. 2011), we will take into account the stricter constraint coming from CMB power spectrum and try to get some further insight on the probability of observing cosmic strings through gravitational lensing events. We will consider ideal strings, that is, cosmic strings approximately straight on the length $L$ of the step of the random walks that they form. 

The Atacama Cosmology Telescope (Dunkley et al. 2011) and Planck data (P. A. R. Ade et al. 2013) provide for NG strings respectively $G\mu/c^{2} < 1.6 \times 10^{-7}$ and $G\mu/c^{2} < 1.5 \times 10^{-7}$. For AH cosmic string networks the limit $G\mu/c^{2} < 3.2 \times 10^{-7}$ is found in the Planck data. According to the above  stringent constraints, let us consider what happens for static strings when the ideal condition of perpendicularity between string orientation and line of sight  is dropped, that is, when $\gamma_{s}(1 + \hat{k} \cdot \vec{\beta_{s}}) |\hat{k} \times \hat{s}|= sin\theta$ and  the angular separation between the lensed images is given by:

\begin{equation} \label{eqn:32}
\delta \phi = \Delta \frac{D_{ls}}{D_{os}} sin \theta    \qquad   \theta \in \left[0,\pi \right].
\end{equation}

\noindent
Since the lensing angle depends on $\Delta$ as well as on the ratio between the angular diameter distances from source to string (lens), $D_{ls}$, and from source to observer, $D_{os}$, and on the angle $\theta$ between the string direction and the line of sight, not all the strings contained in the sphere of the optical sources can be observed through their gravitational lensing effects. 
For a straight string segment lying at $z$ close to the upper limit of the optical range the corresponding lensing angle would be on the border of observability ($\lesssim 0.05^{\prime \prime}$), being $D_{ls} \ll D_{os}$, even in the ideal case $sin \theta = 1$, although $N(z)$ is still quite large for NG strings compared to the AH case being, for instance, in a $\Lambda$CDM cosmology, respectively $\sim 250$ and $\sim 50$. For the most distant sources at $z_s \sim 6-7$, therefore, the string-lens is required to be at $z \lesssim 5$ in order to yield observable effects. For non ideal strings an additional factor less than unity for $\theta \neq \pi/2$ restricts the strings that have more probability to be observed to those located at redshifts around $z \lesssim 2$ (where on average the number of string segments is expected to be $< 55$ for the NG and $< 12$ for the AH model). This implies a cut for the lens location on the highest values of $z$ in the optical sources range  at redshifts where the expected number of string segments, at least in the NG case, is sufficiently  large. Setting a lower bound to the possible redshifts where one can reasonably suppose to find more than one string across the observed sky at the redshift of the transition matter-dark energy, results in a further reduction of the probability of observing gravitational lensing events for the large number of sources (quasars and galaxies) distributed around $z_s \sim 2$ as well as for the most distant and rare sources.

If we consider that the most part of the observed quasars and galaxies are distributed at redshifts $1 \lesssim z_s \lesssim 2$ (Conselice 2004; Stott et al. 2013 and references therein), the range of redshifts at which the probability of observing cosmic strings by their gravitational lensing effects is greatest will be $ 0.6 \leq z < 1$.  For a source at $z_{s} \in [1,2]$ and a segment of string at a redshift $z \in [0.6,1)$, up to values of $z$ whose proximity to the upper limit is model dependent, the corresponding lensing angle is in principle always sufficiently large for ideal strings to produce two distinguishable images of the source when this one is a quasar. As regards galaxies, only the points interior to the strip defined by the deficit angle will appear duplicated the other side of the string. The outer points will be cut away resulting in visible sharp edges in the isophotes of the source images (Sazhin et al. 2007). However, for $z < 1$  N(z) is less than 15 for NG strings and no more than 3 for AH strings in a $\Lambda$CDM cosmological background. As for quintessence and phantom field models, the  lower and higher number of segments at such $z$ is significant only for NG models. 

Even choosing  a value of the NG string tension parameter very close to the upper bound coming from the Planck data, the deficit angle $\Delta$ will be quite small, being no more than  $0.77^{\prime\prime}$. Consequently, for cosmic strings whose orientation do not match the ideal condition, though located close the lower limit of the matter-dominated era, for some values of $\theta$, the corresponding lensing angle can be too small to be observable. Let us consider, for instance, randomly oriented static strings located at redshifts $0.6 \leq z < 1$ which could be gravitational lenses for any sources in the above mentioned redshifts range. Outside of a certain interval of values of $\theta$, $\delta \phi$ becomes smaller than the angular resolution of current space-based telescopes ($0.1^{\prime \prime}$  to $0.05^{\prime \prime}$). For any considered source, this interval is essentially dependent on the string redshift. Thus, in order to have an idea of the number of potentially observable string segments for any value of that range of $z$ and for any value of the source redshift $z_{s} \in [1,2]$ we have to calculate the probability of observing a segment of cosmic string, in a given volume, which forms  with the line of sight an angle $\theta$ contained in the corresponding interval for each interval separately. By adding the results we find that the total number of segments of static straight string which can produce observable gravitational lensing effects for sources located at $z_{s} \in [1,2]$, depending on the angular resolution of the telescope, for all the considered cosmological background, is reduced of an amount that ranges from about 1/5 (for an angular resolution of $0.05^{\prime \prime}$) to half (for an angular resolution of  $0.1^{\prime \prime}$) of the expected number of string segments at $z \in [0.6,1)$.  
For AH cosmic strings the constraint on the tension parameter set by the Planck data is less strong and consequently the deficit angle can reach  $1.7^{\prime \prime}$, meaning that, although for $ z \in [0.6, 1)$ for any cosmological model is always $N(z) < 4$, almost all of these string segments, however far from the ideal condition in many cases, can give rise to detectable lensing angles.
\\

In Table 5 the second column $N^{obs}$, represents the number of string segments which can  produce a detectable $\delta \phi$ with respect to the number of segments expected in the spherical shell under consideration. In particular, the results relative to string segments outside and inside the source region are separately shown in the subsequent lines of each row "model" of the table.  In the range of the string redshift outside the one chosen for sources, all the string segments in the shell determined by $z \in [0.6, 0.99]$ will give detectable $\delta \phi$  in case of  ideal strings. This number is reduced by an amount dependent primarily on the model describing cosmic strings and to a lesser extent on the cosmological model. The values of the tension parameter chosen for calculation are $1.4 \times 10^{-7}$ and $3.1 \times 10^{-7}$ for classical NG and AH strings, respectively, and $4.4 \times 10^{-7}$ for the string theory $D$ strings. 

Noticing that $N^{obs}$ does not provide the probability of observing the effects of gravitational lensing by cosmic strings, that cannot be taken into account irrespective of the distribution of sources, but only the probability of having $\delta \phi$ sufficiently large to be detected, the range of values of the redshift to which actually may be associated a higher probability of observation is that for which $\delta \phi$ is detectable for any $z_s$. The corresponding number of segments that produce such $\delta \phi$s will be somewhat lower than that given in the first line of the table, relative to the interval $[0.6,1)$, and regards string segments having $z$ that varies in a model-dependent narrower range, $\tilde{z}$. As we can see in the second line of the corresponding row "model"  of Table 5, the about $80\%$ of the segments that can yield observable lensing angles for the NG models and about $ 90\% $ for the AH models are further reduced, on average, to less than $65\%$ and around $80\%$, respectively,if we consider string-source distances such that the segments can be observed as gravitational  lenses for all sources in the range of redshifts $z_s \in [1,2]$. Of course, even for an ideal orientation of the strings there will be a reduction in the number of observable objects due to a lens-source distance too close. The range of $z$ for the segments of string, $\tilde{z}$, that can be observed by means of lensing effects for all $z_s$ is almost the same for both the ideal and non ideal case but the number of objects potentially observable is quite different. For example, for NG strings in a $\Lambda$CDM cosmological background, this difference amounts to $\sim 77\%$ for strings with ideal orientation compared to the $\sim 63\%$ reported in the table for randomly oriented strings.

\begin{center}
\begin{table*}
\caption{$N^{obs}(z,\theta < \pi/2$)   is the probability of observing string segments, forming an angle $\theta$ with respect to the line of sight,  for sources distributed at $z_{s} \in [1,2]$ for different models. $\Delta z$ is the range of the string segments redshifts that define the spherical shells considered for the lensing angle observations. $ \theta < \pi/2: \delta \phi = 0.05^{\prime\prime}$  refers to the largest and smallest deviations from the ideal case such that the lensing angle is detectable for $z_{s}\in [1,2]$ when an angular resolution $0.05^{\prime\prime}$ can be reached. $\delta\phi(\theta = \pi/2)$ refers to the maximum value taken by the lensing angle in the ideal case corresponding to the largest distance string-source for all sources in the  interval  $z_{s} $.}\label{tab:table_5}
\resizebox*{0.85\textwidth}{!}{
\begin{tabular}{|cccccc|}
\hline
\hline  
&&&&&\\ 
Model                      &  $N^{obs}(z,\theta < \pi/2$)   &  $\Delta z$   &  $\theta< \pi/2$ : $\delta\phi = 0.05^{\prime\prime}$   &  $\delta\phi$$ (\theta = \pi/2)$    &    \\
\hline
&&&&&\\
      &  $81\%$ $z_s \in [1,2]$ &   $ z \in [0.6, 1 )$     & $7\pi/180  - 17\pi/36$ &   $0.25^{\prime\prime} - 0.42^{\prime\prime}$      &\\
P (NG)   &$64\%$  $\forall z_s \in [1,2]$ & $\tilde{z} \in [0.6,0.91] $ &  &&\\

&$35\%$ & $ z \in [1, 2)$  &$\pi/15 - 7\pi/15$ &$0.9^{\prime\prime} -  0.26^{\prime\prime}$&\\
&$z_s \in (1,2]$ & $z^{obs} \in [1, 1.76] $ & &   &\\ 
\hline
&&&&&\\
       & $91\%$  $z_s \in [1,2]$ & $ z \in [0.6, 1 )$   &   $\pi/45 - 11\pi/36$   &   $0.55^{\prime\prime} - 0.93^{\prime\prime}$   &\\
P (AH)   & $81\%$  $\forall z_s \in [1,2]$  & $\tilde{z}\in [0.6,0.95] $ & & &\\ 

&  $60\%$& $ z \in [1, 2)$ & $ \pi/36 - 17\pi/45 $  &$0.11^{\prime\prime} - 0.58^{\prime\prime}$&\\ 
& $z_s \in (1,2]$ & $z^{obs} \in [1, 1.88] $  && &\\
\hline
&&&&&\\
    & $80\%$ $z_s \in [1,2]$ &   $ z \in [0.6, 1 )$     &   $7\pi/180 - 22\pi/45$  & $0.24^{\prime\prime} - 0.42^{\prime\prime}$  & \\
$\Lambda$ (NG)  &$63\%$  $\forall z_s \in [1,2]$ &  $\tilde{z} \in [0.6, 0.91]$  &&     &\\

&$36\%$& $ z \in [1, 2)$ &$\pi/15 - 22\pi/45$& $0.09^{\prime\prime} - 0.27^{\prime\prime} $&\\ 
&$z_s \in (1,2]$& $z^{obs} \in [1, 1.77] $  & & &\\ 
\hline
&&&&&\\
 & $91\%$  $z_s \in [1,2]$&  $z \in [0.6, 1) $   &  $\pi/45 - 14\pi/45$ & $0.54^{\prime\prime} - 0.93^{\prime\prime}$  &    \\
 $\Lambda$ (AH)   & $81\%$   $\forall z_s \in [1,2]$   &$ \tilde{z} \in [0.6, 0.95]$&  & & \\

&  $61\%$&$ z \in [1, 2)$& $ \pi/36 - 7\pi/15$ &$0.11^{\prime\prime} - 0.59^{\prime\prime} $ &\\ 
& $z_s \in (1,2]$&$z^{obs} \in [1, 1.89] $ & & &\\ 
\hline
&&&&&\\
  &  $79\%$  $z_s \in [1,2]$ & $z \in [0.6, 1) $   & $2\pi/45 - 4\pi/9$   & $0.23^{\prime\prime} - 0.41^{\prime\prime} $  & \\
 Q (NG)       & $58\%$  $\forall z_s \in [1,2]$ &  $\tilde{z} \in [0.6, 0.9]$ & &  &\\

&$35\%$&  $ z \in [1, 2)$& $\pi/15 - 4\pi/9$& $0.085^{\prime\prime} - 0.25^{\prime\prime}$&\\ 
&$z_s \in (1,2]$& $z^{obs} \in [1, 1.76] $&&&\\
\hline
&&&&&\\
   & $90\%$  $z_s \in [1,2]$  &   $z \in [0.6, 1) $  &   $\pi/45 - 13\pi/36$   & $0.51^{\prime\prime} - 0.9^{\prime\prime} $ & \\
 Q (AH)    & $80\%$  $\forall z_s \in [1,2]$  & $\tilde{z} \in [0.6,0.95] $& &   &\\

&$60\%$& $ z \in [1, 2)$ &$\pi/30 - 7\pi/15$&$0.1^{\prime\prime} - 0.57^{\prime\prime}$&\\ 
& $z_s \in (1,2]$& $z^{obs} \in [1, 1.89] $ & & &\\
\hline
&&&&&\\
    & $93\%$  $z_s \in [1,2]$ &   $z \in [0.6, 1) $     &   $\pi/60 - 4\pi/15$   & $0.77^{\prime\prime} - 1.3^{\prime\prime}$& \\
D strings  & $85\%$   $\forall z_s \in [1,2]$  &$ \tilde{z} \in [0.6, 0.96]$& &  &\\

$g_s = 0.04$&$70\%$&$ z \in [1, 2)$ & $\pi/45 - 2\pi/5$& $0.15^{\prime\prime} - 0.84^{\prime\prime}$&\\ 
&$z_s \in (1,2]$& $z^{obs} \in [1, 1.92] $ &&&\\ 
\hline
\end{tabular}
}
\end{table*}
\end{center}

In Table 5  it is also shown the maximum value of $\delta \phi$ when $\theta= \pi/2$ for sources between $z_s=1$ and $z_s=2$. The lowest value of $\delta \phi$, assumed at the smallest distances string-source, is taken to be the angular resolution $0.05^{\prime \prime}$. How small this distance can be depends more consistently on the cosmic string model than the cosmological model as well as the largest value that $\delta \phi$ can assume in correspondence of the largest distance lens-source also reported in the Table. This last column shows how the two quantities  $r = D_{ls} / D_{os}$ and  $\Delta$ in  eq. (\ref{eqn:32}), related respectively to the cosmological model and the cosmic string model, affect the value of  $\delta \phi$ and what are the maximum values that can be reached in the best possible situation of the mutual distances string-source-observer. To such best situation is associated a large range of the allowed deviation from the ideal case given by the first number of the $\theta$ column. For instance, for NG strings in a $\Lambda $CDM cosmology, $\theta$ can vary from $7^{o}$ to $90^{o}$ with detectable values of $\delta \phi$ ranging from $0.05^{\prime \prime}$ to the interval of values given in the last column. In the fourth column, the values of the maximum and minimum deviation $\theta$ such that the lensing angle be observable, in function of the distance string-source, show that when $D_{ls}$ becomes very small, we can have $\delta \phi$ observable only for $\theta$ values very close to $90^{o}$,  i.e., only for strings nearly perpendicular to the line of sight. For instance, for a NG model in a $\Lambda$CDM cosmological background, $\delta \phi$ is detectable for $ z$ of the lens very close to $z_s = 1$, $z = 0.91$, only if  $\theta \geq 88^{o}$, i. e.,  for strings with almost-ideal orientation, while for $P$  models a greater deviation from the ideal condition is possible, and the same distance in a $Q$ model does not provide  a detectable $\delta \phi$ even in the ideal case. A very small allowed deviation from the perpendicularity between string and line of sight is confined to the case of very small lens-source distances. In the majority of cases significant deviations from the ideal case can be considered,  meaning that realistic cosmic strings may have detectable gravitational lensing effects to an extent that is not too far from the upper bound exhibited in Table 5. Regarding the AH strings, for any cosmological model, very close distances lens-source as well as  quite large deviations from the ideal case, may produce a detectable $\delta \phi$. This is also true for $ D$ strings. However, whatever the considered model and  distance lens-source, the range of  $\theta$ for which  $\delta \phi$ is detectable does never include all values of $\theta$. The largest interval is obtained for $D$ strings:  $\delta \phi \geq 0.05^{\prime \prime}$ for $\theta \geq 3^{o}$ 
\\

When the lens is at a redshift  $1\leq z < 2$, within the volume considered for the sources, as $z$ increases, the ratio between the angular distances  $D_{ls}$ and $D_{os}$ decreases  so that even small deviations from the ideal case can yield undetectable lensing angles. For NG strings  $\delta \phi$ is no longer detectable  for  $z \sim 1.8$ even for cosmic strings with ideal orientation and assuming an angular resolution of $0.05^{\prime \prime}$. Of all the string segments contained in the spherical shell determined by $z \in [1,2)$ only a small fraction may have a $\delta \phi$ potentially observable for sources at  $z_{s} \in (1,2]$. For the  AH models, although the expected number of string segments at $z \in [1,2)$  ranges from $ 2-4$ to $6-15$, since the deficit angle is large enough, a large number of them ($\sim 60\%$) can produce a $\delta \phi$ detectable  up to $z \sim 1.9 $, for sources at $z_{s} \in (1,2]$, for some values of $\theta$ of the randomly oriented strings.  In Table 5,  $z^{obs}$  provides the range of redshifts for string segments within the region of sources to have detectable $\delta \phi$ when some deviation from the ideal condition is taken into account. 

As it can be seen in the above mentioned Table, $\Lambda $CDM  and $P$  models show very little difference in the percentage of objects that can produce detectable lensing effects. In this interval of the string redshift, apart from the number of possible lenses certainly not very high, it is also crucial the reduction of the number of possible sources as $z$ increases.

The small increase in $N^{obs}$ resulting in the case of a $\Lambda $CDM model  compared to a $P$ model case, apparently in contrast with what happens in the range of $z$ of the string-lens outside the region where the sources are located, is explained by the different trend of the ratio $r = D_{ls} / D_{os}$ which, for  strings at $z \in [1,2)$, for growing values of the redshift of the string  approaching the source close to the upper bound of the considered region, reverses its behavior by passing from $r_P> r_{\Lambda}$ to $r_{\Lambda}> r_P$, while this never occurs for strings at $z \in [0.6,1)$. For quintessence NG and AH models, being always $r_Q < r_{\Lambda}$ and only for the lens location close to the source at $z_s$ near the upper limit $r_Q > r_P$, $N^{obs}$ in this range of the string redshift becomes comparable to that obtained for NG and AH phantom fields models.  

$N^{obs}$ shows some difference  among  cosmic strings models and little difference among cosmological models as regards the reduction due to deviations from the ideal case. This was expected, since the different value of $\Delta$ affects the results more than the difference in the ratio $r$. As a consequence, being  the number of expected segments  different for different cosmological models, we will end up with a different number of objects observable through gravitational lensing effects. Evidently, the worse situation is  given by the Q models.  At this respect it is worth mentioning that Planck observations show that CMB data are compatible with a cosmological constant, as assumed in the  $\Lambda$CDM model. Whereas additional SNLS SNe data favor the phantom domain $w < -1$  (P. A. R. Ade et al. 2013 and references therein). 
\\

It is worth pointing out that, according to the representation suggested by the scaling solution, a network of random walk strings can be approximated by a collection of straight string segments, of length given by the step of the random walk, fairly uniformly distributed. The string segments have random orientation and move with randomly directed velocities. Since strings are relativistic, at each Hubble time a different distribution of cosmic string segments affects the photons propagating freely after last scattering or emitted at later times by galaxies. We consider the centers of the string segments located at different redshift to estimate the approximate number of string segments inside our backward light cone. At any time $t_{*}$  before $t_{0}$, considering the set of all events in the past light cone of the observer at  $t_{0}$ which occur at that time $t_{*}$, the result would be a sphere centered on the present observer with a radius  given by the distance traveled by photons from that time to today. Any point on or inside the sphere could send a signal moving at the speed of light or slower that would have time to reach the observer at $t_{0}$, while points outside the sphere at that moment would not be able to have any causal contact with the present observer. Since we wish to detect cosmic strings not through direct emission but by means of photon worldlines interconnected with the string worldsheets, in order to find the effective number of segments which produce deflection and frequency shift of photons, the light cone projection of the string segments needs to be considered (see e.g. Hindmarsh 1994).

An exhaustive treatment on the topic involving survey fields, distribution of sources at different redshifts and their brightness, the string velocity and effects due to the wiggliness is beyond the scope of this paper. Here we only mean to point out  to what extent the underlying models, either for the cosmological background and for cosmic strings,  can make the difference between having or not observable effects. To show this we focused on the most favorable redshift range of sources, that is, around the star formation peak where the observed objects are more numerous and brighter. 
\\

The outcome of the above discussion is in large part a consequence of the small value of the deficit angle $\Delta$, as we have shown. Thus, when we apply the same considerations to cosmic superstrings we find significant differences  due to the fact that for a small value of the string coupling constant ($g_{s}\sim 10^{-2}$), while the $D$ and $FD$ components of the network  have tensions comparable to those of ordinary cosmic strings, the $F$ component is expected to have smaller tension, $G\mu_{F}/c^{2} \leq 1.8 \times 10^{-8}$. This leads to very small values of the lensing angle, $\delta \phi < 0.05^{\prime \prime}$, which prevents any optical identification even in the ideal case of string segments perpendicular to the line of sight. As a consequence, although a small value of the string coupling constant yields an exceedingly large number of string segments ($\sim 10^{3}$) for the $F$ component even at small redshifts (down to the value $z \sim 0.6$) against the small number found for ordinary cosmic strings, models supporting a small value of the string coupling constant cannot be ruled out on the grounds of the predicted large number of the $F$ component as long as the range of small values of tension involved are still unexplored by means of current technology. We will not dwell on the case $g_{s} = 0.9$, considering the relative results only by way of comparison with the interesting opposite  case of small values of the string coupling constant.  In fact, models with a small $g_{s}$, which ensure a compactification scale exceeding or equal to the four-dimensional GUT scale and lead to a low-energy supergravity, allow also to work in a perturbative regime consistently with the underlying theory. A small value of the deficit angle such as the one obtained for the $F$ component in the  limit $g_{s}\sim 10^{-2}$,  $\Delta \sim 0.09^{\prime \prime}$,  leads to undetectable lensing angles. On the other hand, as regards the $D$ component of the network, being  $G\mu_{D}/c^{2} \leq 4.5 \times 10^{-7}$, the deficit angle approaches the value $2.3^{\prime \prime}$, thus leading to detectable gravitational  lensing effects for about the $80\%$ of the randomly oriented segments of  $D$ strings in the entire volume considered. Cosmic superstrings can produce detectable $\delta \phi $ up to very short string-source distances and with a greater allowed deviation from the ideal condition for observation. 

Although $\mu_{FD} \simeq \mu_{D}$, the very small number of the $FD$ strings, leads to the conclusion that if for $D$ strings, as well as for traditional strings,  gravitational  lensing is expected to be a very rare event, this is even more true in the case of the Y-junctions that may form from $FD$ bound states.
\\

\section{Acknowledgements}
The authors would like to thank Cosimo Stornaiolo for helpful discussions.
The research was financially supported by the grant RFFI 10-02-00961a. The work was carried out as part of the project No. 14.740.11.0085 of the Ministry of Education. RC acknowledges partial financial support by the NASA - Jet Propulsion Laboratory and the kind hospitality of the California Institute of Technology, where this work was partly done.

\end{document}